\documentclass{emulateapj}

\usepackage{apjfonts, natbib}

\newcommand{\be}{\begin{equation}}
\newcommand{\ee}{\end{equation}}

\shorttitle{First results from QUaD}
\shortauthors{QUaD collaboration}

\begin{document}

\slugcomment{Submitted to ApJ}

\title{First season QUaD CMB temperature and polarization power spectra}

\author{
  QUaD collaboration
  --
  P.\,Ade\altaffilmark{1},
  J.\,Bock\altaffilmark{2,3},
  M.\,Bowden\altaffilmark{1,4},
  M.\,L.\,Brown\altaffilmark{5,6},
  G.\,Cahill\altaffilmark{7},
  J.\,E.\,Carlstrom\altaffilmark{8},
  P.\,G.\,Castro\altaffilmark{5,9},
  S.\,Church\altaffilmark{4},
  T.\,Culverhouse\altaffilmark{8},
  R.\,Friedman\altaffilmark{8},
  K.\,Ganga\altaffilmark{10},
  W.\,K.\,Gear\altaffilmark{1},
  J.\,Hinderks\altaffilmark{4,11},
  J.\,Kovac\altaffilmark{3},
  A.\,E.\,Lange\altaffilmark{3},
  E.\,Leitch\altaffilmark{2,3},
  S.\,J.\,Melhuish\altaffilmark{1,12},
  J.\,A.\,Murphy\altaffilmark{7},
  A.\,Orlando\altaffilmark{1},
  R.\,Schwarz\altaffilmark{8},
  C.\,O'\,Sullivan\altaffilmark{7},
  L.\,Piccirillo\altaffilmark{1,12},
  C.\,Pryke\altaffilmark{8},
  N.\,Rajguru\altaffilmark{1,13},
  B.\,Rusholme\altaffilmark{4,14},
  A.\,N.\,Taylor\altaffilmark{5},
  K.\,L.\,Thompson\altaffilmark{4},
  E.\,Y.\,S.\,Wu\altaffilmark{4}
  and
  M.\,Zemcov\altaffilmark{1,2,3}
}

\altaffiltext{1}{School of Physics and Astronomy, Cardiff University,
  Queen's Buildings, The Parade, Cardiff CF24 3AA, UK.}
\altaffiltext{2}{Jet Propulsion Laboratory, 4800 Oak Grove Dr.,
  Pasadena, CA 91109, USA.}
\altaffiltext{3}{California Institute of Technology, Pasadena, CA
  91125, USA.}
\altaffiltext{4}{Kavli Institute for Particle Astrophysics and
Cosmology, Stanford University, 382 Via Pueblo Mall,
  Stanford, CA 94305, USA.}
\altaffiltext{5}{Institute for Astronomy, University of Edinburgh,
  Royal Observatory, Blackford Hill, Edinburgh EH9 3HJ, UK.}
\altaffiltext{6}{{\em Current address}: Cavendish Laboratory,
  University of Cambridge, J.J. Thomson Avenue, Cambridge CB3 OHE, UK.}
\altaffiltext{7}{Department of Experimental Physics,
  National University of Ireland Maynooth, Maynooth, Co. Kildare,
  Ireland.}
\altaffiltext{8}{Kavli Institute for Cosmological Physics,
  Department of Astronomy \& Astrophysics, Enrico Fermi Institute, University of Chicago,
  5640 South Ellis Avenue, Chicago, IL 60637, USA.}
\altaffiltext{9}{{\em Current address}: CENTRA, Departamento de F\'{\i}sica,
  Edif\'{\i}cio Ci\^{e}ncia, Piso 4,
  Instituto Superior T\'ecnico - IST, Universidade T\'ecnica de Lisboa,
  Av. Rovisco Pais 1, 1049-001 Lisboa, Portugal.}
\altaffiltext{10}{Laboratoire APC/CNRS, B\^atiment Condorcet,
  10, rue Alice Domon et L\'eonie Duquet, 75205 Paris Cedex 13, France.}
\altaffiltext{11}{{\em Current address}: NASA Goddard Space Flight
  Center, 8800 Greenbelt Road, Greenbelt, Maryland 20771, USA.}
\altaffiltext{12}{{\em Current address}: School of Physics and
  Astronomy, University of
  Manchester, Manchester M13 9PL, UK.}
\altaffiltext{13}{Department of Physics and Astronomy, University
  College London, Gower Street, London WC1E 6BT, UK.}
\altaffiltext{14}{{\em Current address}:
  Infrared Processing and Analysis Center,
  California Institute of Technology, Pasadena, CA 91125, USA.}

\begin{abstract}
QUaD is a bolometric CMB polarimeter sited at the South Pole,
operating at frequencies of 100 and 150~GHz.
In this paper we report preliminary results from the first season of
operation (austral winter 2005).
All six CMB power spectra are presented derived as cross spectra
between the 100 and 150~GHz maps using 67 days of observation
in a low foreground region of approximately 60~square degrees.
This data is a small fraction of the data acquired to date.
The measured spectra are consistent with the $\Lambda$CDM cosmological model.
We perform jackknife tests which indicate that the observed signal
has negligible contamination from instrumental systematics.
In addition by using a frequency jackknife we find no evidence for
foreground contamination.
\end{abstract}


\section{Introduction}
The CMB is expected to be polarized at the $\sim10\%$ level due to
Thomson scattering by free electrons of the local quadrupole in the
CMB radiation field at the time of last scattering.
The resulting polarization signal can be decomposed into two independent modes.
At the time of last scattering, even parity $E$-modes are generated by both
scalar and tensor (gravitational wave) metric perturbations while
odd-parity $B$-modes are generated only by gravitational waves.
A secondary source of $B$-mode polarization comes from the weak
gravitational lensing effect of intervening large scale structure,
which converts $E$-modes into $B$-modes on small scales.

The first detection of the $E$-mode polarization signal was made by the
30~GHz radio interferometer, DASI, in 2002~\citep{kovac02}.
Since then, in addition to a further measurement
by the DASI experiment~\citep{leitch05}, $E$-mode
measurements have been made with the CBI~\citep{readhead04},
CAPMAP~\citep{barkats05}, BOOMERanG~\citep{montroy06},
WMAP~\citep{page06} and MAXIPOL~\citep{wu06} experiments.

High precision measurements of the $E$-mode signal represent a
non-trivial test of the standard cosmological model since the
polarization of the CMB probes the velocity field at the time of last
scattering, as opposed to the density field probed by CMB temperature
measurements.
In addition, accurate measurements of $E$-mode polarization
can be useful for constraining certain cosmological
parameters which are fairly insensitive to the CMB temperature field
(e.g. isocurvature modes in the early Universe).
Such a high resolution measurement of the $E$-mode polarization signal is the
\setcounter{footnote}{0}
primary science goal of the QUaD\footnote{
QUaD stands for ``QUEST and DASI''.
In turn, QUEST is ``Q \& U Extragalactic Survey Telescope'' and DASI
stands for ``Degree Angular Scale Interferometer''.
The two experiments merged to become QUaD in 2003.}
experiment.
In addition to the polarization, QUaD will, with further analysis,
also provide interesting results on the CMB temperature field on small
scales.

In this paper we present preliminary power spectra measured from
QUaD's first season of operation.
The paper is organized as follows.
In Section~\ref{sec:inst}, we summarize the QUaD instrument
and describe the observation strategy and low-level data reduction.
Section~\ref{sec:maps} describes our map-making and simulation procedure.
Our power spectrum estimation method is described in Section~\ref{sec:spectra_method}
and the power spectrum results are presented
in Section~\ref{sec:spectra_results} along with results from a number of
jackknife tests.
In Section~\ref{sec:param_estimation} we estimate cosmological parameters
using our spectra and our conclusions are presented
in Section~\ref{sec:conclusions}.

Throughout this paper when we refer to ``the $\Lambda$CDM model''
we mean specifically the model generated by the CMBFAST
program~\citep{zaldarriaga00} using the WMAP3 cosmological parameters given under the heading
``Three Year Mean'' in table 2 of \cite{spergel06}.
This is the model used in our simulations and shown in the plots.

\section{Instrument Summary and Observations}
\label{sec:inst}

QUaD is a millimeter-wavelength bolometric polarimeter designed for
observing the CMB at two frequency bands, 100 and 150~GHz.
The experiment is sited at the MAPO observatory, approximately 1~km from
the geographic South Pole.
First light was achieved in February 2005, and science observations began in May 2005.
The telescope is a 2.6~m on-axis Cassegrain with nominal beam sizes of 6.3
(100~GHz) and 4.2 (150~GHz) arcminutes.
The tower, ground shield and altitude-azimuth mount of the DASI experiment
are re-used for QUaD, the ground shield being extended to accommodate
the larger telescope structure.
The mount has a third axis which rotates
about the optical symmetry axis (termed ``deck'' rotation).
This is a very useful feature for a polarimeter, as it allows the entire
telescope to be rotated to an arbitrary angle with respect to the
sky.

The QUaD receiver comprises two anti-reflection coated cryogenic
re-imaging lenses and a focal plane array of 31~pixels, each composed
of a corrugated feed horn and two orthogonal Polarization-Sensitive
Bolometers (PSBs;~\cite{2003SPIE.4855..227J}).
The PSB pairs are oriented on the focal plane in two groups with bolometer
sensitivity angles separated by 45~degrees.
This redundancy in detector orientation allows one to construct
maps of the sky in Stokes $Q$ \& $U$ with observations at a single
deck angle if so desired.
Each pixel is single-frequency and the pixels are divided
between the two observing bands with 12 at 100~GHz and 19 at 150~GHz.
The PSBs are similar to those flown on the successful B2K experiment~\citep{masi06}.
A complete description of the receiver along with details of the optical
testing and characterization will be provided in Hinderks et al. (in prep).

The first season of QUaD observations were completed in October 2005 and
consisted of $\sim$100 days of CMB runs, in addition to special runs
for pointing model determination, beam mapping, and detector time constant
measurements.
The CMB runs consist of scanning the telescope back and forth
by 7.5 degrees in azimuth, in a series of 30 second constant-elevation ``half-scans''.
The telescope is then stepped by 0.02 degrees in elevation and the process repeated
to build up a raster map.
Since the telescope is sited close to the Earth's axis of rotation,
azimuth and elevation closely approximate to right ascension
and declination.
In the first season we have mapped a 60 square degree patch in the
low-foreground B2K deep region~\citep{masi06}.
For our chosen scanning speed and observing declination $\ell \sim 2000 f$,
where $f$ is the frequency in Hz at which multipole number $\ell$ appears in the
time ordered data.
The timeconstants of most (80\%) of our detectors are less than 30~ms
with the slowest two $\sim 100$~ms corresponding to half power
roll-offs at $\ell \sim 10000$ and
$\ell \sim 3000$ respectively.

To permit the removal of ground contamination, the scanning
strategy employs a ``lead-trail'' scheme whereby each hour of
observations is split equally between a ``lead'' field (1st half hour)
and a ``trail'' field (2nd half hour), separated by 0.5 hours in right ascension.
The lead and trail field observations follow exactly the same pattern in
telescope azimuth and elevation so a constant ground signal can be
removed by differencing the lead and trail field data.
Furthermore, each day of observation is split into two 8 hour blocks made
over different ranges in azimuth with the telescope rotated at different deck angles.
This enables a powerful jackknife test as described below.

Initial processing of the raw time-ordered data (TOD) consists of glitch removal,
deconvolution of the bolometer temporal response, low pass filtering and down-sampling.
The relative calibration factor between channels (and within channels
of a given pair) is derived from frequent short scans in elevation
(el-nods) which introduce a strong atmospheric gradient into the TOD.
This relative calibration is applied separately within each frequency
group.
Various quality control data cuts are applied at this stage:
days with bad weather, bad pointing, poor focal plane temperature stability,
or moon contamination are discarded.
After applying these data cuts, 67 of the $\sim$100 days of CMB observations remain,
and are used in the following science analysis.

\section{Map-making and Simulation Process}
\label{sec:maps}

Two analysis pipelines have been constructed which are independent in the
sense that they share no code, although the algorithms are intended to
be identical (with some important exceptions --- see below).
For this initial analysis we use data which has been point by point
lead-trail differenced as described above.
There is clear ground pickup in the data, which although mostly common mode,
has a polarized component.
This pickup appears to be completely canceled in the field difference.
We note that many CMB experiments have mitigated ground pickup by
field differencing, for example DASI.

Before mapping, a best-fit 3rd order
polynomial is subtracted from each half-scan to remove the bulk of the
$1/f$ noise.  The signal from each PSB pair is then summed
and differenced and co-added into the map according to the telescope
pointing information.
In the co-addition, a weighting is applied
according to the inverse variance of each 30 second half-scan
to properly account for differences
in sensitivity across pairs, and to down weight periods of
poorer weather.
The pair sum data is used to construct total intensity ($T$) maps
and the pair difference data is used to construct
maps of Stokes $Q$ and $U$ by using the known orientation angle of each
detector pair with respect to the sky.
During this process, a $\sim 10\%$ correction for the non-ideal
polarization efficiency is applied to the differenced PSB data.
These angles and efficiencies are measured
from special observations of a chopped, polarized, thermal source
placed externally to the telescope.

While building the $T$, $Q$ and $U$ maps, we
also construct the expected $T$, $Q$ and $U$ variance maps under
the simple assumption that the noise is uncorrelated (white) in the TOD.
These variance maps are used later to weight the signal maps in
the power spectrum estimation stage.

As described in the next section the power spectrum estimation
technique requires signal only, noise only,
and signal plus noise simulations of the complete experiment.
We generate these as follows.

To construct simulated noise timestreams,
we Fourier transform each set of half-scans and take
the auto and cross spectra between all channel pairs.
This allows us to re-generate simulated noise timestreams
with the same frequency spectra as the real data {\it and}
the proper cross correlations between channels.
We assume that the instantaneous signal to noise in the time-ordered 
data is sufficiently low that the power spectra created in this way 
accurately estimate the noise.

To generate simulated signal timestreams we generate realizations of
$T$, $Q$ and $U$ sky maps under
the $\Lambda$CDM model using the ``synfast'' generator
(part of the HEALPix package\footnote
{See http://healpix.jpl.nasa.gov/index.shtml and \cite{gorski05}}),
convolve with the instrumental beam (separately for each detector),
and re-sample according to the pointing of the telescope.
Included in this process are a scatter in the polarization
efficiency factors of the PSB pairs and a scatter in the PSB sensitivity
angles.
The uncertainties used are derived from the special observations of a chopped,
polarized thermal source mentioned above.
A detailed beam model is used which is derived from special beam mapping runs on
the compact HII region RCW38, daily scans of each detector across
this source, and observations of a bright quasar PMN J0538-4405 which
lies within our field.

Either of the signal or noise simulated timestream, or their sum,
is then passed through the standard
mapping algorithm (complete with polynomial subtraction and variance
weighting) to yield simulated maps.

To derive the absolute calibration factor of our experiment we pass
the B2K temperature maps~\citep{masi06} through the simulation process
to provide maps which are filtered in exactly the same way as the QUaD
maps, and are thus directly comparable to them.  Both sets of maps are
then Fourier transformed and cross spectra taken between them to
determine the relative calibration factor.  B2K is in turn calibrated
against WMAP.
(WMAP3 lacks sufficient sensitivity within our small sky area
to allow a direct cross calibration.)
We perform this calibration separately for the 100~GHz
and 150~GHz maps, resulting in $5\%$ absolute calibration
uncertainty in units of temperature ($10\%$ in units of power),
including both QUaD and BOOMERanG
uncertainties~\citep{masi06}.
To monitor the absolute stability of the
instrument an internal calibration source is inserted into
the beam frequently during routine data taking and these readings show excellent
stability over the entire season of observations (few percent for
any given channel and $0.5\%$ for the gain ratio of a PSB pair).

\begin{figure}
  \epsscale{1.2}
  \plotone{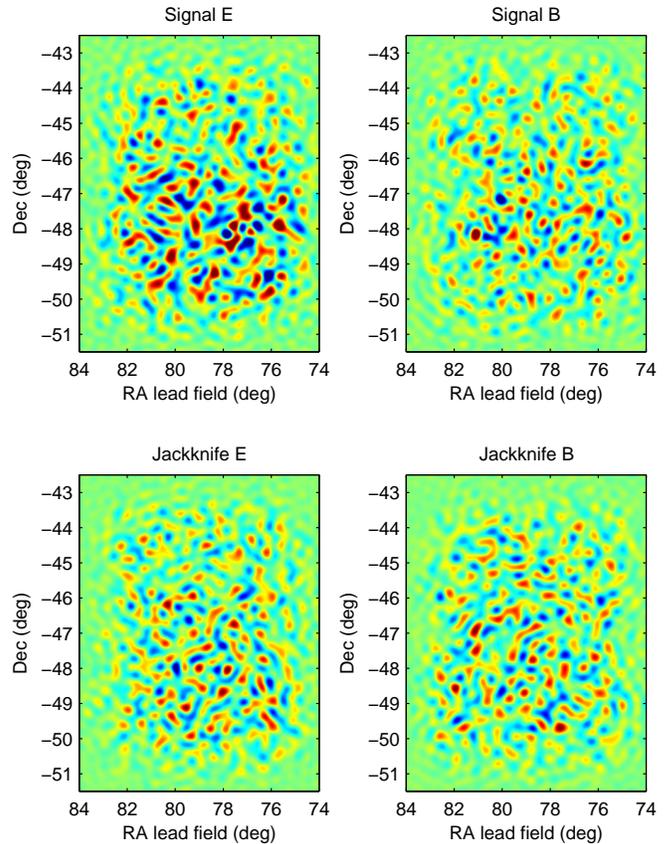}
  \caption{QUaD first season field differenced polarization maps decomposed into $E$ and $B$-modes,
and filtered to include only the angular scale range $200<\ell<1000$.
The top row shows the result for signal (non-jackknife) maps, while
the bottom row is for the ``deck'' jackknife (see text).
This plot shows 150~GHz maps and the color scale is $\pm30$~$\mu$K$^2$ in all cases.}
  \label{fig:maps}
\end{figure}

For visual presentation only, we divide by the variance map,
transform to $E$ and $B$, and filter to the angular scales where signal
to noise is highest, to produce Figure~\ref{fig:maps}.

\section{Power Spectrum Estimation Method}
\label{sec:spectra_method}

To estimate angular power spectra, we employ a Monte Carlo (MC)
based analysis.
This method requires the creation of noise only simulated
power spectra to correct the measured power spectra for noise bias,
and signal only power spectra to allow the suppression
of power by filtering to be corrected.
In addition signal plus noise spectra are required to provide
the final covariance matrix of the bandpower measurements.

Before measuring power spectra from the real and simulated maps, we
apply an inverse variance mask to the maps based on the expected
spatial distribution of the noise (using the variance maps mentioned above),
and additionally mask a small number (5) of point sources which are apparent
in our $T$ maps, and confirmed by external catalogs.
We note that none of these sources is detected at high significance
in the $Q$ and $U$ maps.

At this point the two pipelines diverge --- one
follows the standard MASTER technique of \cite{hivon02}, extended to
polarization \citep{brown05}, and works explicitly
on the curved sky (pseudo-$C_\ell$), while the other makes the flat sky
approximation and uses two dimensional Fourier transforms to derive
power spectra.

The first pipeline measures raw pseudo-$C_\ell$ power spectra from the maps
using a modified version of the ``anafast'' program included in the
HEALPix package.
Estimates of the CMB power spectra are then reconstructed as bandpowers
(${\bf P}_b$) from the pseudo-$C_\ell$ measurements using
\begin{equation}
{\bf P}_b=\sum_{b'}{\bf M}^{-1}_{bb'} \, \sum_\ell P_{b'\ell} \,
(\widetilde{\bf C}_\ell-\langle \widetilde{\bf N}_\ell \rangle_{MC}).
\label{bandpowers}
\end{equation}
Here, $P_{b\ell}$ is a binning operator in $\ell$-space,
$\widetilde{\bf C}_\ell$ are the raw pseudo-$C_\ell$ spectra
measured from the real data and $\langle \widetilde{\bf N}_\ell
\rangle_{MC}$ are the average pseudo spectra measured from the
noise only simulations.
${\bf M}_{bb'}$ is the binned coupling matrix of \cite{brown05}
which corrects for mode-mixing due to the survey
geometry.
${\bf M}_{bb'}$ also contains the correction for the effects of
both the TOD polynomial filtering and the telescope beam width.
These corrections are derived from the set of signal only simulations.
Finally, the covariance matrix of the resulting bandpowers is found from
the scatter among the power spectra measured from simulations containing
signal and noise.

The second pipeline takes the two dimensional Fourier
transform of the masked maps,
converts the $Q$ and $U$ Fourier modes into $E$ and $B$ and
calculates bandpowers as the mean of the product of the modes
within each annular bin.
The product can be taken as the auto spectrum of a given
map, or as a cross spectrum between two maps (which need not
be at the same frequency).
This is done for the real data maps and for each simulation realization.
The data spectra then have the mean of the corresponding set of noise only
simulations subtracted to noise correct them.
Filter/beam suppression factors are calculated as the ratio of the
mean of the signal only simulations to the input power spectra,
and the data spectra are corrected by dividing out these
suppression factors.
Finally the bandpower covariance matrix is estimated from the scatter of
the signal plus noise simulations in the same way as for
pipeline one.

The first pipeline explicitly corrects for mixing between the $EE$
and $BB$ spectra due to the sky cut using the ${\bf M}_{bb'}$ matrix.
The second pipeline does not make such a correction, but the level
of mixing into the $BB$ spectra in the signal only simulations
is found to be negligible compared to the current
instrumental sensitivity ($<0.2$~$\mu$K$^2$).
In addition note that the simulations indicate that inter- and
intra-spectra mode mixing due to the half-scan polynomial filtering, and lack of
cross linking in the map, are absolutely irrelevant for the current
analysis.

For either pipeline we can take spectra internally within the sets
of 100 and 150~GHz maps, or cross spectra between the two frequencies.
In this paper, we present only the frequency cross spectra.
The single frequency spectra will be presented in a future paper.

\section{Results and Jackknife Tests}
\label{sec:spectra_results}

\begin{figure}
\epsscale{1.2}
\plotone{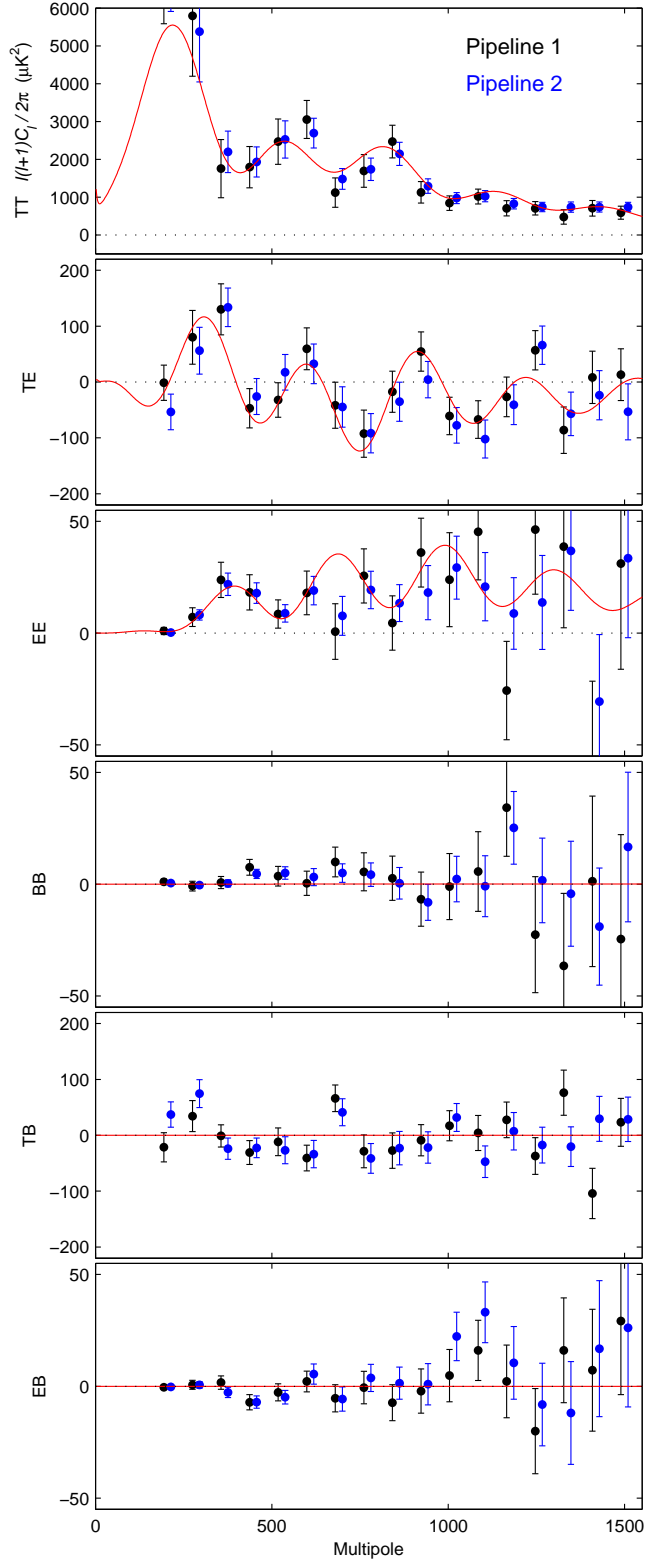}
\caption{Preliminary QUaD first season results
derived as cross power spectra between the 100 and 150~GHz maps.
The two sets of points are the results from two completely independent
analyzes of the data (see text for details) and are displaced
by $\pm10$ in multipole number from their nominal values
for clarity.}
\label{fig:spec1}
\end{figure}

\begin{figure}
\includegraphics[scale=0.75, angle=-90.]{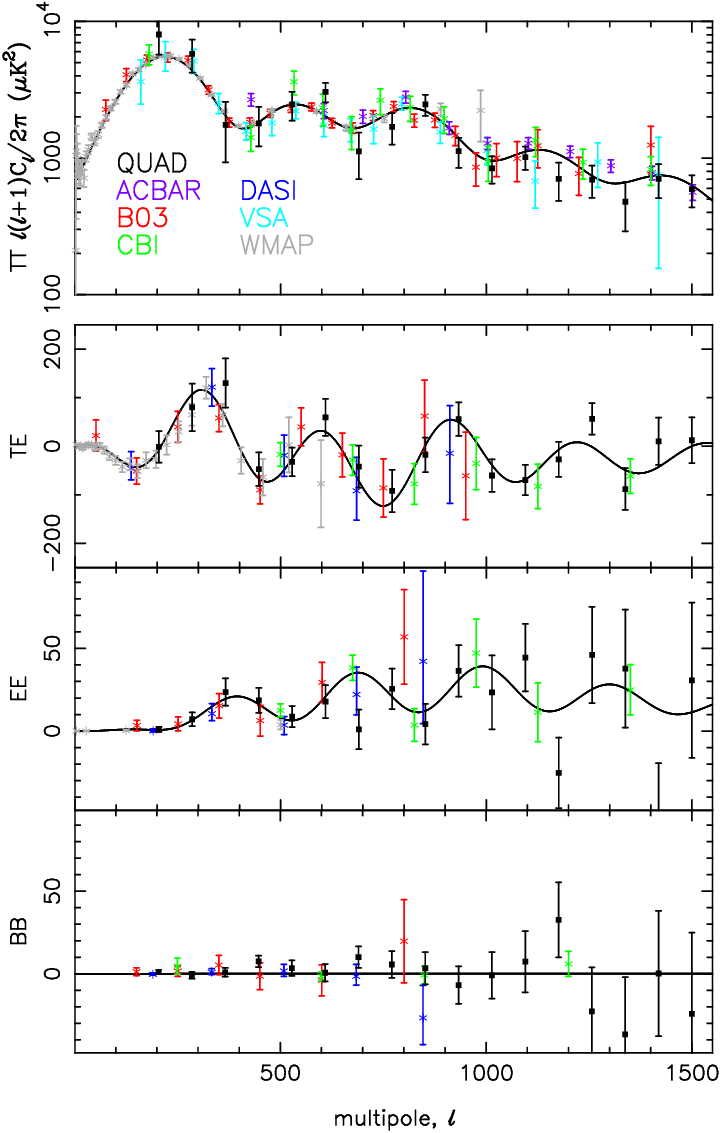}
\vspace{3mm}
\caption{Comparison of preliminary QUaD first season power
spectrum results to selected other published results to date. The
experiments compared to are ACBAR \citep{kuo06}, BOOMERanG
\citep{jones06, piacentini06, montroy06}, CBI \citep{sievers05}, DASI
\citep{leitch05}, VSA \citep{dickinson04} and WMAP \citep{hinshaw06,
  page06}. Note that the $TT$ comparison is plotted with a log scale
in the y-axis.}
\label{fig:spec2}
\end{figure}

Figure~\ref{fig:spec1} shows the frequency cross spectra measured from
the first season of QUaD data by the two independent pipelines\footnote
{Bandpowers, covariance matrices and bandpower window functions are
available in numerical form at http://quad.uchicago.edu/quad}.
The error bars are the square root of the diagonal elements of the bandpower
covariance matrices which are estimated from the run-to-run scatter among
MC simulations as described in the previous section.
The first pipeline includes a mode decoupling
step which narrows the $\ell$ range to which each bandpower
responds, while at the same time changing the adjacent bandpower correlation
coefficients from their ``natural'' value of $\approx+0.2$ to $\approx-0.2$.
This leads to an increase in the diagonal of the bandpower
covariance matrix which is reflected by larger error bars on the
plot.
However we emphasize that the total information content of the
bandpowers from both pipelines is similar.
In Figure~\ref{fig:spec2} our results are shown compared to
published results from other experiments.

A powerful test for systematic contamination from ground pickup, or
another source, is the so-called jackknife test.
In this paper we use a map based jackknife, forming separate maps
from various approximately equally sized data subsets, subtracting these
and taking the power spectra of the result.
We also do this for the signal plus noise simulations to
estimate the expected uncertainty of the jackknife spectra.
In as much as the signal originates on the sky it should
exactly cancel under jackknife --- depending on the split,
and its origin, false signal is not likely to do so.
Here we use the following data splits each of which
will be explained in turn: deck split, scan direction
split, season split, focal plane split and frequency split.

\begin{figure}
\epsscale{1.2}
\plotone{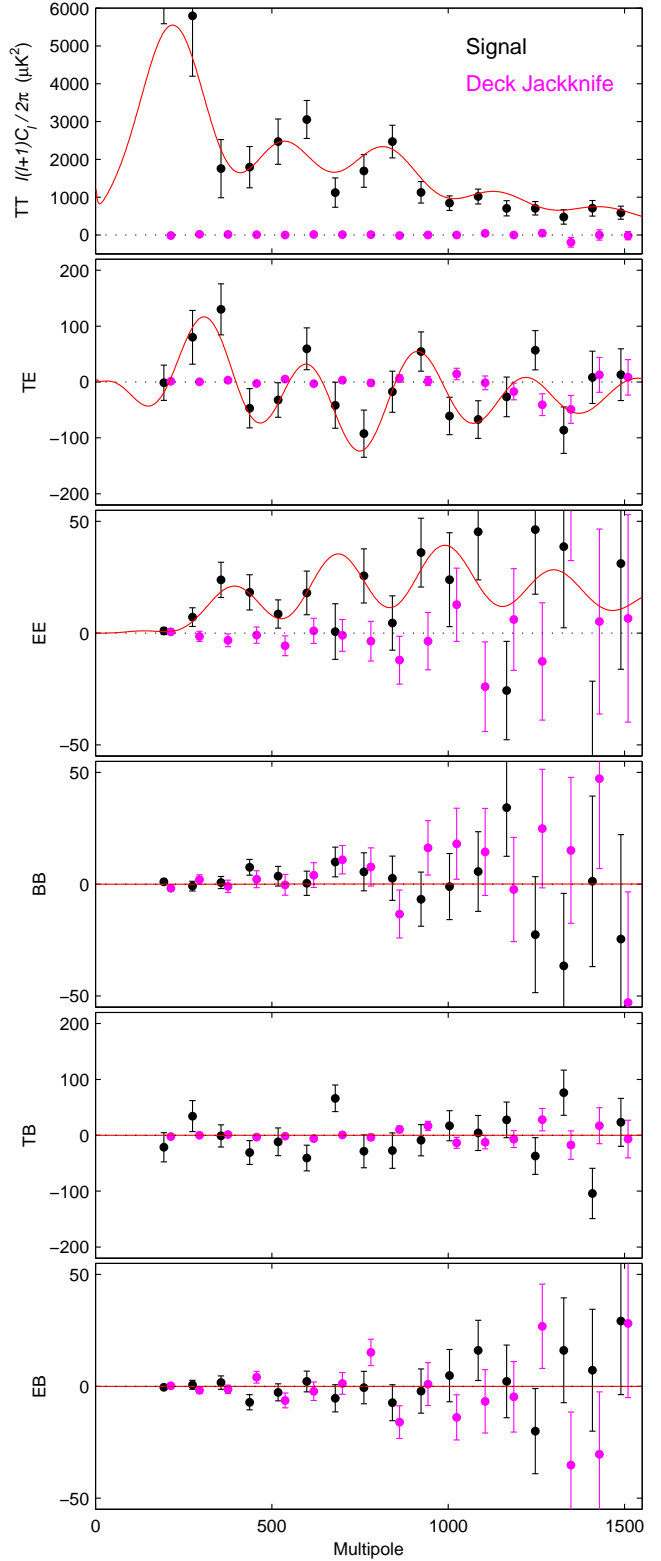}
\caption{QUaD signal spectra compared to deck jackknife spectra.
(See text for details.)}
\label{fig:dkjack}
\end{figure}

The so called deck split is possibly the most powerful test.
As mentioned above each day of observations on our CMB field
is split into two 8 hour blocks.
Because the run starts always at the same local sidereal
time these blocks occur always over the same range of azimuth
angle as the telescope turns within its ground shield.
In addition each block of observations is made always at the same
rotation angle of the telescope with respect to the line of sight,
with a 60 degree rotation occurring between the two blocks.
Hence each given bolometer pair scans the sky at a different
orientation angle within each block.
Therefore the deck jackknife polarization
maps will only cancel if the rotation to the absolute reference
frame which occurs in the map making step is being performed
correctly.
The ground pickup is observed to be very complex with certain
pair differences showing a detectable spike always at a certain azimuth angle
and rotation angle of the telescope.
Hence even if some ground pickup is leaking through the field
differencing operation, we certainly do not expect it to
appear identically in the deck split $Q$ and $U$ maps
and thus to cancel in the deck jackknife.
Figure~\ref{fig:dkjack} compares the signal (non-jackknife)
and deck jackknife power spectra and we see that
the vast majority of the apparent sky signal cancels.
Note that where the signal spectra are sample variance dominated the
jackknife spectra error bars are smaller since there is no sample
variance in a null spectrum.

For the scan direction jackknife we form separate maps from
the half-scans in each direction.
If deconvolution of the detector temporal response is not done
correctly then the forward and backward maps will not match and
residuals will remain when they are subtracted.
In some cases the time constants of the two halves of a detector
pair are not well matched and hence poor deconvolution
could lead to leakage from $T$ into polarization making this
a very important test.

The split season jackknife forms maps from the first half and second
half of the used days.
If for example there were a drift in the absolute calibration of the instrument
over time then cancellation failure would be expected here.

The focal plane split forms separate maps using the two orientation
groups of bolometer pairs in the focal plane.
Because observations are taken at two deck angles it is possible
to construct $Q$ and $U$ maps using each group.

\begin{figure}
\epsscale{1.2}
\plotone{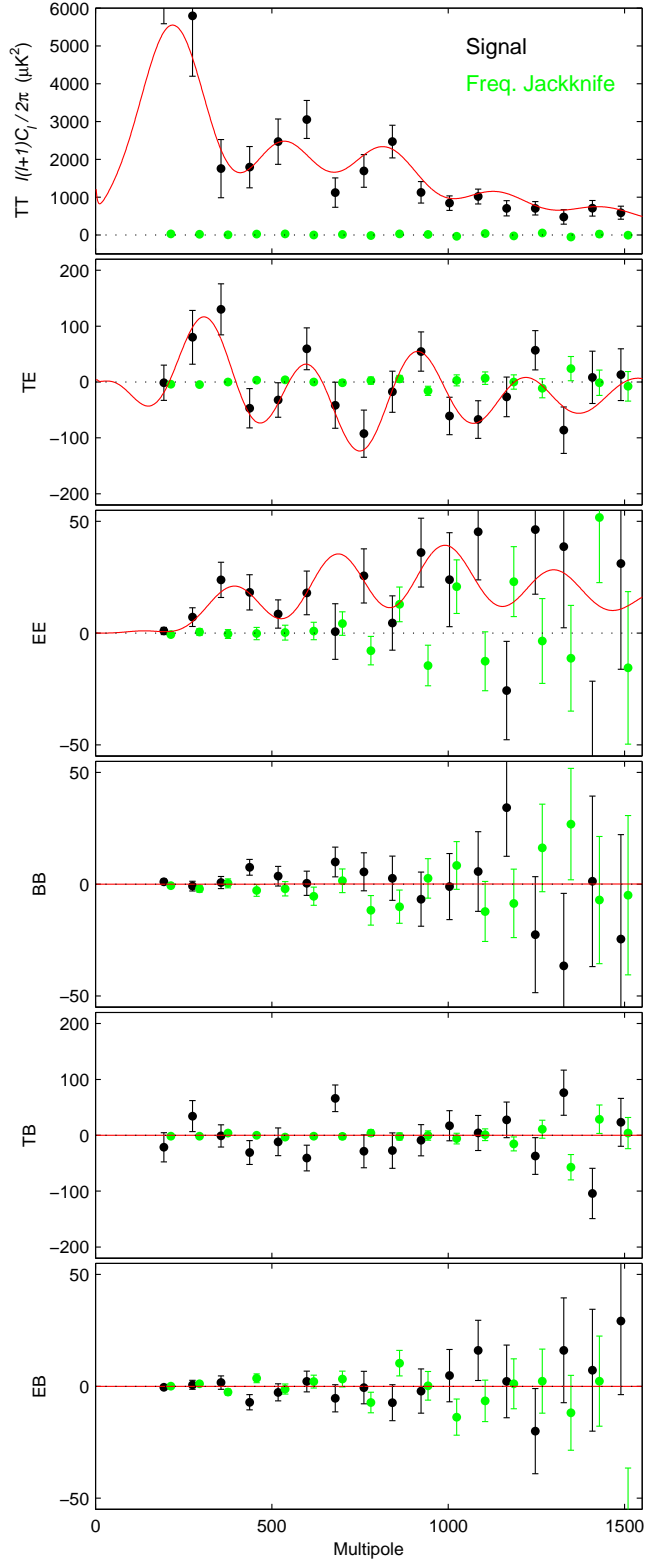}
\caption{QUaD signal spectra compared to frequency jackknife spectra.
(See text for details.)}
\label{fig:fqjack}
\end{figure}

For the frequency jackknife instead of taking cross spectra between the
100~GHz and 150~GHz maps we subtract them and take the
spectra of the frequency difference maps.
Any admixture of two or more signal components with differing
spatial distributions and frequency spectral indexes is expected to fail
this test (for example CMB plus synchrotron and/or dust) ---
this is therefore a stringent test for foreground contamination.
Figure~\ref{fig:fqjack} compares the signal (non-jackknife)
and frequency jackknife power spectra and we see that to the limits of
experimental sensitivity the sky pattern is identical at the
two frequencies.

Figures~\ref{fig:dkjack} and~\ref{fig:fqjack} are visually
impressive but to quantify how well these tests are passed we have
calculated $\chi^2$ statistics for the comparison of the
jackknife power spectra with the null model.
For some of the jackknifes we do not expect perfect cancellation
due to the interaction of the polynomial filtering by the half-scan and
the non perfectly overlapping coverage region of the two data subsets.
Hence we compare the measured $\chi^2$ values to the distribution which
we measure from the set of signal plus noise simulations rather than to
a theoretical $\chi^2$ distribution.

In Table~\ref{tab:chi2}, the probability to exceed (PTE) the $\chi^2$
value measured from the data is tabulated for each spectrum and
jackknife test.
These numbers are for the first pipeline --- the second pipeline
gives results which are similar.
Ideally these PTE values should be uniformly distributed from zero to one.
The only value which shows an obvious problem is
$TT$ in the scan direction case --- the enormous signal to noise of
the $TT$ spectrum seen in Figures~\ref{fig:dkjack} and~\ref{fig:fqjack}
makes the $TT$ jackknifes sensitive to tiny systematic errors.
The jackknife tests of the single frequency spectra show some additional
problems passing these formal $\chi^2$ tests and we are hence choosing not
to publish these spectra at this time.

The final row of Table~\ref{tab:chi2} is not a jackknife --- it is a comparison
of our measured spectra against the WMAP3 $\Lambda$CDM model mentioned
above and shown in the plots.
The PTE values lie within the acceptable range
indicating that our measured spectra are consistent with this model.

\begin{deluxetable}{c c c c c c c}
  \tablewidth{8cm} \tablecaption{PTE values from $\chi^2$
  tests\label{tab:chi2}} \tablehead{\colhead{Jackknife} &
  \colhead{$TT$} & \colhead{$TE$} & \colhead{$EE$} & \colhead{$BB$} &
  \colhead{$TB$} &\colhead{$EB$}}
    \startdata
    Deck angle &               0.236 & 0.208 & 0.812 & 0.435 & 0.274 & 0.062 \\
    Scan direction &           0.000 & 0.173 & 0.304 & 0.375 & 0.236 & 0.223 \\
    Split season &             0.032 & 0.814 & 0.257 & 0.527 & 0.904 & 0.111 \\
    Focal plane &              0.193 & 0.702 & 0.079 & 0.503 & 0.450 & 0.225 \\
    Frequency &                0.034 & 0.306 & 0.610 & 0.452 & 0.642 & 0.135 \\
    Signal\tablenotemark{a} &  0.501 & 0.964 & 0.415 & 0.482 & 0.066 & 0.809 \\
    \enddata
  \tablenotetext{a}{The PTE value for the signal case is
  calculated against the $\Lambda$CDM model.}
\end{deluxetable}

\section{Cosmological Parameter Estimation}
\label{sec:param_estimation}

We have carried out a basic 6 parameter cosmological parameter constraint
analysis using our polarization power spectra only, i.e. we use the
$TE$, $EE$ and $BB$ spectra, but not $TT$.
Our methodology uses the Monte-Carlo Markov Chain (MCMC) method and
is based on that of the WMAP team as described in \cite{verde03};
we impose the same flat priors, the same re-parameterization and the same
convergence/mixing test.
In addition we optimize the Markov chain step size choice,
a fundamental parameter for the correct behavior of the algorithm,
by calculating the parameter covariance matrix of a preliminary run of our 
chains as in \cite{tegmark04}.
We deal with the beam and calibration uncertainty using a 
method proposed by \cite{bridle02} which accomplishes an effective
marginalization over these parameters by adding extra terms to the
bandpower covariance matrix.

The theoretical power spectra are calculated using CAMB \citep{lewis00}, 
and then transformed by means of the experimental bandpower window functions 
to predictions for the binned $P_b$'s.
We assume that the likelihood distribution for our bandpowers is
Gaussian, which our signal plus noise simulations indicate is a valid assumption.
The bandpower covariance matrix for the $P_b$'s is assumed to be 
independent of the model, and is estimated from the signal plus noise
simulations as described above.

In Table \ref{tab:param} we present the marginalized expectation values and
68\% confidence limits for each parameter.
Four converged MCMC chains with around 50,000 steps each are merged to obtain these
results.
For all of the parameters our 68\% confidence limit encloses the WMAP3
expectation value.
The $\chi^2$ of the model with the parameters listed in the table is 44.1 giving a probability
to exceed this by chance of 0.51 (for the 45 degrees of freedom).

\begin{deluxetable}{l c c }
  \tablewidth{5cm} 
  \tablecaption{
	Polarization only cosmological parameter constraints using preliminary
	QUaD first season power spectra
\label{tab:param}}
\tablehead{\colhead{Parameter} & \colhead{Symbol} & \colhead{Value}}
    \startdata
    Baryon density              & $\Omega_bh^2$    &        $0.0260\,\,_{-0.0059}^{+0.0061}$    \\
    Matter density              & $\Omega_mh^2$    &        $0.142\,\,_{-0.031}^{+0.029}$ \\
    Hubble constant             &      $h$        &        $0.74\,\,_{-0.15}^{+0.16}$    \\
    Optical depth               &     $\tau$      &        $0.156\,\,_{-0.088}^{+0.089}$ \\
    Scalar fluctuation amplitude\tablenotemark{a} &     $A_s$       &        $0.78\,\,_{-0.16}^{+0.16}$    \\
    Scalar fluctuation index\tablenotemark{a}    &      $n_s$      &        $0.833\,\,_{-0.293}^{+0.283}$ \\
    \enddata
    \tablenotetext{a}{The pivot point for $A_s$ and $n_s$ is $k_p=0.05$~Mpc$^{-1}$}
\end{deluxetable}

\section{Conclusions}
\label{sec:conclusions}

We have presented preliminary power spectra measured from the first
season of observations with QUaD.
In this paper we have presented only the
frequency cross spectra taken between the 100 and 150~GHz maps.
We find that these spectra
are entirely consistent with the $\Lambda$CDM model ---
the measured $EE$ spectrum has a distinctive peak at $\ell\sim400$
exactly as expected, the $TE$ spectrum shows the
expected correlations, and the $BB$ spectrum is consistent with zero.
A basic polarization only parameter constraint analysis yields
confidence limits which agree with the WMAP3 results, and
are as tight as those which were derived from CMB temperature
spectra just a few years ago.

We have performed jackknife tests by measuring power spectra from differenced
maps generated under several data splits and find that the results
are free from significant instrumental systematics.
In addition we have presented a frequency jackknife which indicates
that contamination of the CMB by astrophysical foregrounds
is negligible for the current experimental sensitivity.

We note that this analysis considers only frequency cross spectra
and includes only 67 days out of a total of 250 days of CMB
observing to date.

The QUaD experiment has begun a third season of observations
and analysis of the second season data is underway. When completed, we
fully expect to improve substantially on the preliminary results
presented here.

\acknowledgements

QUaD is funded by the National Science Foundation in the USA, through
grants AST-0096778, ANT-0338138, ANT-0338335 \& ANT-0338238, by the
Particle Physics and Astronomy Research Council in the UK and by the
Science Foundation Ireland. We would like to thank the staff of the
Amundsen-Scott South Pole Station and all involved in the United States
Antarctic Program for the superb support operation which makes the
science presented here possible. Special thanks go to our intrepid
winter over scientist Robert Schwarz who has spent three consecutive
winter seasons tending the QUaD experiment. MLB acknowledges the award
of a PPARC fellowship. SEC acknowledges support from a Stanford Terman
Fellowship. JRH acknowledges the support of an NSF Graduate Research
Fellowship and a Stanford Graduate Fellowship. CP and JEC acknowledge
partial support from the Kavli Institute for Cosmological Physics
through the grant NSF PHY-0114422.
EYW acknowledges receipt of an NDSEG fellowship.

\bibliographystyle{apj}
\bibliography{ms}

\begin{thebibliography}{24}
\expandafter\ifx\csname natexlab\endcsname\relax\def\natexlab#1{#1}\fi

\bibitem[{{Barkats} {et~al.}(2005){Barkats}, {Bischoff}, {Farese},
  {Fitzpatrick}, {Gaier}, {Gundersen}, {Hedman}, {Hyatt}, {McMahon},
  {Samtleben}, {Staggs}, {Vanderlinde}, \& {Winstein}}]{barkats05}
{Barkats}, D. {et~al.} 2005, \apjl, 619, L127, astro-ph/0409380

\bibitem[{{Bridle} {et~al.}(2002){Bridle}, {Crittenden}, {Melchiorri},
  {Hobson}, {Kneissl}, \& {Lasenby}}]{bridle02}
{Bridle}, S.~L., {Crittenden}, R., {Melchiorri}, A., {Hobson}, M.~P.,
  {Kneissl}, R., \& {Lasenby}, A.~N. 2002, \mnras, 335, 1193, astro-ph/0112114

\bibitem[{{Brown} {et~al.}(2005){Brown}, {Castro}, \& {Taylor}}]{brown05}
{Brown}, M.~L., {Castro}, P.~G., \& {Taylor}, A.~N. 2005, \mnras, 360, 1262,
  astro-ph/0410394

\bibitem[{{Dickinson} {et~al.}(2004){Dickinson}, {Battye}, {Carreira},
  {Cleary}, {Davies}, {Davis}, {Genova-Santos}, {Grainge}, {Guti{\'e}rrez},
  {Hafez}, {Hobson}, {Jones}, {Kneissl}, {Lancaster}, {Lasenby}, {Leahy},
  {Maisinger}, {{\"O}dman}, {Pooley}, {Rajguru}, {Rebolo}, {Rubi{\~n}o-Martin},
  {Saunders}, {Savage}, {Scaife}, {Scott}, {Slosar}, {Sosa Molina}, {Taylor},
  {Titterington}, {Waldram}, {Watson}, \& {Wilkinson}}]{dickinson04}
{Dickinson}, C. {et~al.} 2004, \mnras, 353, 732, arXiv:astro-ph/0402498

\bibitem[{{G{\'o}rski} {et~al.}(2005){G{\'o}rski}, {Hivon}, {Banday},
  {Wandelt}, {Hansen}, {Reinecke}, \& {Bartelmann}}]{gorski05}
{G{\'o}rski}, K.~M., {Hivon}, E., {Banday}, A.~J., {Wandelt}, B.~D., {Hansen},
  F.~K., {Reinecke}, M., \& {Bartelmann}, M. 2005, \apj, 622, 759,
  astro-ph/0409513

\bibitem[{{Hinshaw} {et~al.}(2006){Hinshaw}, {Nolta}, {Bennett}, {Bean},
  {Dor{\'e}}, {Greason}, {Halpern}, {Hill}, {Jarosik}, {Kogut}, {Komatsu},
  {Limon}, {Odegard}, {Meyer}, {Page}, {Peiris}, {Spergel}, {Tucker}, {Verde},
  {Weiland}, {Wollack}, \& {Wright}}]{hinshaw06}
{Hinshaw}, G. {et~al.} 2006, ArXiv Astrophysics e-prints, astro-ph/0603451

\bibitem[{{Hivon} {et~al.}(2002){Hivon}, {G{\'o}rski}, {Netterfield}, {Crill},
  {Prunet}, \& {Hansen}}]{hivon02}
{Hivon}, E., {G{\'o}rski}, K.~M., {Netterfield}, C.~B., {Crill}, B.~P.,
  {Prunet}, S., \& {Hansen}, F. 2002, \apj, 567, 2, astro-ph/0105302

\bibitem[{{Jones} {et~al.}(2006){Jones}, {Ade}, {Bock}, {Bond}, {Borrill},
  {Boscaleri}, {Cabella}, {Contaldi}, {Crill}, {de Bernardis}, {De Gasperis},
  {de Oliveira-Costa}, {De Troia}, {di Stefano}, {Hivon}, {Jaffe}, {Kisner},
  {Lange}, {MacTavish}, {Masi}, {Mauskopf}, {Melchiorri}, {Montroy}, {Natoli},
  {Netterfield}, {Pascale}, {Piacentini}, {Pogosyan}, {Polenta}, {Prunet},
  {Ricciardi}, {Romeo}, {Ruhl}, {Santini}, {Tegmark}, {Veneziani}, \&
  {Vittorio}}]{jones06}
{Jones}, W.~C. {et~al.} 2006, \apj, 647, 823, arXiv:astro-ph/0507494

\bibitem[{{Jones} {et~al.}(2003){Jones}, {Bhatia}, {Bock}, \&
  {Lange}}]{2003SPIE.4855..227J}
{Jones}, W.~C., {Bhatia}, R., {Bock}, J.~J., \& {Lange}, A.~E. 2003, in
  Presented at the Society of Photo-Optical Instrumentation Engineers (SPIE)
  Conference, Vol. 4855, Millimeter and Submillimeter Detectors for Astronomy.
  Edited by Phillips, Thomas G.; Zmuidzinas, Jonas. Proceedings of the SPIE,
  Volume 4855, pp. 227-238 (2003)., ed. T.~G. {Phillips} \& J.~{Zmuidzinas},
  227--238

\bibitem[{{Kovac} {et~al.}(2002){Kovac}, {Leitch}, {Pryke}, {Carlstrom},
  {Halverson}, \& {Holzapfel}}]{kovac02}
{Kovac}, J.~M., {Leitch}, E.~M., {Pryke}, C., {Carlstrom}, J.~E., {Halverson},
  N.~W., \& {Holzapfel}, W.~L. 2002, Nature, 420, 772, astro-ph/0209478

\bibitem[{{Kuo} {et~al.}(2006){Kuo}, {Ade}, {Bock}, {Bond}, {Contaldi}, {Daub},
  {Goldstein}, {Holzapfel}, {Lange}, {Lueker}, {Newcomb}, {Peterson},
  {Reichardt}, {Ruhl}, {Runyan}, \& {Staniszweski}}]{kuo06}
{Kuo}, C.~L. {et~al.} 2006, ArXiv Astrophysics e-prints, astro-ph/0611198

\bibitem[{{Leitch} {et~al.}(2005){Leitch}, {Kovac}, {Halverson}, {Carlstrom},
  {Pryke}, \& {Smith}}]{leitch05}
{Leitch}, E.~M., {Kovac}, J.~M., {Halverson}, N.~W., {Carlstrom}, J.~E.,
  {Pryke}, C., \& {Smith}, M.~W.~E. 2005, \apj, 624, 10, astro-ph/0409357

\bibitem[{{Lewis} {et~al.}(2000){Lewis}, {Challinor}, \& {Lasenby}}]{lewis00}
{Lewis}, A., {Challinor}, A., \& {Lasenby}, A. 2000, \apj, 538, 473,
  arXiv:astro-ph/9911177

\bibitem[{{Masi} {et~al.}(2006){Masi}, {Ade}, {Bock}, {Bond}, {Borrill},
  {Boscaleri}, {Cabella}, {Contaldi}, {Crill}, {de Bernardis}, {de Gasperis},
  {de Oliveira-Costa}, {de Troia}, {di Stefano}, {Ehlers}, {Hivon}, {Hristov},
  {Iacoangeli}, {Jaffe}, {Jones}, {Kisner}, {Lange}, {MacTavish}, {Marini
  Bettolo}, {Mason}, {Mauskopf}, {Montroy}, {Nati}, {Nati}, {Natoli},
  {Netterfield}, {Pascale}, {Piacentini}, {Pogosyan}, {Polenta}, {Prunet},
  {Ricciardi}, {Romeo}, {Ruhl}, {Santini}, {Tegmark}, {Torbet}, {Veneziani}, \&
  {Vittorio}}]{masi06}
{Masi}, S. {et~al.} 2006, \aap, 458, 687, astro-ph/0507509

\bibitem[{{Montroy} {et~al.}(2006){Montroy}, {Ade}, {Bock}, {Bond}, {Borrill},
  {Boscaleri}, {Cabella}, {Contaldi}, {Crill}, {de Bernardis}, {De Gasperis},
  {de Oliveira-Costa}, {De Troia}, {di Stefano}, {Hivon}, {Jaffe}, {Kisner},
  {Jones}, {Lange}, {Masi}, {Mauskopf}, {MacTavish}, {Melchiorri}, {Natoli},
  {Netterfield}, {Pascale}, {Piacentini}, {Pogosyan}, {Polenta}, {Prunet},
  {Ricciardi}, {Romeo}, {Ruhl}, {Santini}, {Tegmark}, {Veneziani}, \&
  {Vittorio}}]{montroy06}
{Montroy}, T.~E. {et~al.} 2006, \apj, 647, 813, astro-ph/0507514

\bibitem[{{Page} {et~al.}(2006){Page}, {Hinshaw}, {Komatsu}, {Nolta},
  {Spergel}, {Bennett}, {Barnes}, {Bean}, {Dor{\'e}}, {Halpern}, {Hill},
  {Jarosik}, {Kogut}, {Limon}, {Meyer}, {Odegard}, {Peiris}, {Tucker}, {Verde},
  {Weiland}, {Wollack}, \& {Wright}}]{page06}
{Page}, L. {et~al.} 2006, ArXiv Astrophysics e-prints, astro-ph/0603450

\bibitem[{{Piacentini} {et~al.}(2006){Piacentini}, {Ade}, {Bock}, {Bond},
  {Borrill}, {Boscaleri}, {Cabella}, {Contaldi}, {Crill}, {de Bernardis}, {De
  Gasperis}, {de Oliveira-Costa}, {De Troia}, {di Stefano}, {Hivon}, {Jaffe},
  {Kisner}, {Jones}, {Lange}, {Masi}, {Mauskopf}, {MacTavish}, {Melchiorri},
  {Montroy}, {Natoli}, {Netterfield}, {Pascale}, {Pogosyan}, {Polenta},
  {Prunet}, {Ricciardi}, {Romeo}, {Ruhl}, {Santini}, {Tegmark}, {Veneziani}, \&
  {Vittorio}}]{piacentini06}
{Piacentini}, F. {et~al.} 2006, \apj, 647, 833, arXiv:astro-ph/0507507

\bibitem[{{Readhead} {et~al.}(2004){Readhead}, {Myers}, {Pearson}, {Sievers},
  {Mason}, {Contaldi}, {Bond}, {Bustos}, {Altamirano}, {Achermann}, {Bronfman},
  {Carlstrom}, {Cartwright}, {Casassus}, {Dickinson}, {Holzapfel}, {Kovac},
  {Leitch}, {May}, {Padin}, {Pogosyan}, {Pospieszalski}, {Pryke}, {Reeves},
  {Shepherd}, \& {Torres}}]{readhead04}
{Readhead}, A.~C.~S. {et~al.} 2004, Science, 306, 836, astro-ph/0409569

\bibitem[{{Sievers} {et~al.}(2005){Sievers}, {Achermann}, {Bond}, {Bronfman},
  {Bustos}, {Contaldi}, {Dickinson}, {Ferreira}, {Jones}, {Lewis}, {Mason},
  {May}, {Myers}, {Padin}, {Pearson}, {Pospieszalski}, {Readhead}, {Reeves},
  {Taylor}, \& {Torres}}]{sievers05}
{Sievers}, J.~L. {et~al.} 2005, ArXiv Astrophysics e-prints, astro-ph/0509203

\bibitem[{{Spergel} {et~al.}(2006){Spergel}, {Bean}, {Dor{\'e}}, {Nolta},
  {Bennett}, {Hinshaw}, {Jarosik}, {Komatsu}, {Page}, {Peiris}, {Verde},
  {Barnes}, {Halpern}, {Hill}, {Kogut}, {Limon}, {Meyer}, {Odegard}, {Tucker},
  {Weiland}, {Wollack}, \& {Wright}}]{spergel06}
{Spergel}, D.~N. {et~al.} 2006, ArXiv Astrophysics e-prints, astro-ph/0603449

\bibitem[{{Tegmark} {et~al.}(2004){Tegmark}, {Strauss}, {Blanton}, {Abazajian},
  {Dodelson}, {Sandvik}, {Wang}, {Weinberg}, {Zehavi}, {Bahcall}, {Hoyle},
  {Schlegel}, {Scoccimarro}, {Vogeley}, {Berlind}, {Budavari}, {Connolly},
  {Eisenstein}, {Finkbeiner}, {Frieman}, {Gunn}, {Hui}, {Jain}, {Johnston},
  {Kent}, {Lin}, {Nakajima}, {Nichol}, {Ostriker}, {Pope}, {Scranton},
  {Seljak}, {Sheth}, {Stebbins}, {Szalay}, {Szapudi}, {Xu}, {Annis},
  {Brinkmann}, {Burles}, {Castander}, {Csabai}, {Loveday}, {Doi}, {Fukugita},
  {Gillespie}, {Hennessy}, {Hogg}, {Ivezi{\'c}}, {Knapp}, {Lamb}, {Lee},
  {Lupton}, {McKay}, {Kunszt}, {Munn}, {O'Connell}, {Peoples}, {Pier},
  {Richmond}, {Rockosi}, {Schneider}, {Stoughton}, {Tucker}, {vanden Berk},
  {Yanny}, \& {York}}]{tegmark04}
{Tegmark}, M. {et~al.} 2004, \prd, 69, 103501, astro-ph/0310723

\bibitem[{{Verde} {et~al.}(2003){Verde}, {Peiris}, {Spergel}, {Nolta},
  {Bennett}, {Halpern}, {Hinshaw}, {Jarosik}, {Kogut}, {Limon}, {Meyer},
  {Page}, {Tucker}, {Wollack}, \& {Wright}}]{verde03}
{Verde}, L. {et~al.} 2003, \apjs, 148, 195, astro-ph/0302218

\bibitem[{{Wu} {et~al.}(2006){Wu}, {Zuntz}, {Abroe}, {Ade}, {Bock}, {Borrill},
  {Collins}, {Hanany}, {Jaffe}, {Johnson}, {Jones}, {Lee}, {Matsumura},
  {Rabii}, {Renbarger}, {Richards}, {Smoot}, {Stompor}, {Tran}, \&
  {Winant}}]{wu06}
{Wu}, J.~H.~P. {et~al.} 2006, ArXiv Astrophysics e-prints, astro-ph/0611392

\bibitem[{{Zaldarriaga} \& {Seljak}(2000)}]{zaldarriaga00}
{Zaldarriaga}, M., \& {Seljak}, U. 2000, \apjs, 129, 431, astro-ph/9911219

\end{thebibliography}

\end{document}